\documentstyle[]{article}
\font\cero=cmss10 scaled 1728 \font\uno=cmssbx10 scaled 1200
\begin{document}
\begin{flushleft}
{\cero Fluctuating topological invariants in string theory as an
Abelian gauge theory} \\[3em]
\end{flushleft}
{\sf R. Cartas-Fuentevilla}\\
{\it Instituto de F\'{\i}sica, Universidad Aut\'onoma de Puebla,
Apartado postal J-48 72570, Puebla Pue., M\'exico
(rcartas@sirio.ifuap.buap.mx).}  \\[2em]

It is shown that the topological invariants associated with the
two-dimensional world-surface in string theory have nontrivial
fluctuations around their nonexistent classical dynamics.
Additionally it is proved that the underlying geometrical
structure in a covariant phase space formulation for such
topological string actions mimics entirely that of an Abelian
gauge theory.\\

\noindent {\uno I. Introduction} \vspace{1em}

Recently \cite{1,2} it has been demonstrated that the presence of
topological terms in Lagrangians for string theory has a dramatic
effect on the covariant phase space formulation of the theory,
despite such terms do not have an effective contribution on the
equations of motion. This fact is not certainly exclusive of
string theory; for example, in \cite{3} the relevant role that the
topological terms play in the covariant canonical formalism for
4-dimensional BF theory has been explored, and additionally it is
clarified that the knowledge of the equations of motion for a
physical system is not enough for specifying the symplectic
properties of the phase space, but an action principle is a
necessary ingredient for such a purpose. An immediate consequence
of these results is that the presence of topological terms will
lead in general to a completely different quantum field theory.

Traditionally the topological terms have been considered as
corrective or additional terms to other Lagrangian terms, which
normally have nontrivial equations of motion. Subsequently one
will try, for example, to observe the shift on the resultant
quantum field theory. In the present work we attempt to explore an
extreme situation, which was suggested in \cite{2}, and consists
in considering that the only Lagrangian term in a string action is
a topological invariant associated with the 2-dimensional manifold
of the worldsheet. We want to show then that such a topological
string action has, by itself, a nontrivial covariant phase space
formulation, and consequently a nontrivial quantum field theory.
The specific result will be that the underlying symplectic
structure of the topological action mimics  entirely that of an
Abelian gauge theory, with all what such a result might imply.

The philosophy behind this work is that the classical equations of
motion are not the most important thing in physics; in fact, as we
shall see, we do not need, at all, such equations {\it for making
physics}. In some sense, we are not making something new, since it
is known that in a cohomological topological field theory of the
Witten type one tries to make physics from topological invariants
of certain manifolds, which have trivial classical dynamics, but
the physics is found in other domain (see for example \cite{4}).

This work is organized as follows. In the next section we give an
outline of the differential geome\-try of an embedding developed
by Carter \cite{5}, on which we base our calculations. In Section
III we summarize briefly the relevant results of \cite{2} that are
important in the present context. In Section IV we show the
existence of a nontrivial fluctuation dynamics for a topological
invariant, and a covariantly conserved current is constructed from
it, which will be identified with the integral kernel of the
symplectic structure for the topological string action. In Section
V we prove that the integral kernel is obtained also from the
variations of a symplectic potential, following the ideas
discussed previously in \cite{2}. In Section VI we describe the
analogy between the symplectic structure of the topological
invariant and that  of an Abelian gauge theory. In Section VII the
analogy is extended to the symmetry properties of the geometrical
structures. In Section VIII we discuss how to extend the
definition of covariant phase space for topological terms without
classical equations of motion. We finish in Section IX with some
remarks and
prospects.\\

\noindent {\uno II. Basic differential geometry of an imbedding}
\vspace{1em}

In this Section, we outline the description given in \cite{5} for
the intrinsic curvature that is associated with a spacelike or
timelike $p$-surface imbedded in an $n$-dimensional space or
spacetime background with metric $g_{\mu\nu}$. Specifically the
{\it internal curvature tensor} of the imbedding can be written as
\begin{equation}
     R_{\kappa\lambda} {^{\mu}}_{\nu} = 2 n_{\sigma} {^{\mu}} \ n_{\nu}
     {^{\tau}} \ n_{[\lambda} {^{\pi}} \ \overline{\nabla}_{\kappa ]}
     \ \rho_{\pi} {^{\sigma}}_{\tau} + 2 \rho_{[\kappa} {^{\mu\pi}}
     \ \rho_{\lambda ]\pi\nu},
\end{equation}
where $n^{\mu\nu}$ is the (first) {\it fundamental tensor} of the
$p$-surface, that together with the complementary orthogonal
projection $\bot^{\mu\nu}$ satisfy
\begin{equation}
     n^{\mu} {_{\nu}} + \bot^{\mu} {_{\nu}} = g^{\mu} {_{\nu}},
     \quad n^{\mu} {_{\nu}} \ \bot^{\nu} {_{\rho}} = 0,
\end{equation}
and the tangential covariant differentiation operator is defined
in terms of the fundamental tensor as
\begin{equation}
     \overline{\nabla}_{\mu} = n^{\rho} {_{\mu}} \ \nabla_{\rho},
\end{equation}
where $\nabla_{\rho}$ is the usual Riemannian covariant
differentiation operator associated with $g_{\mu\nu}$.
Additionally, $\rho_{\lambda} {^{\mu}}_{\nu}$ represents the
background spacetime components of the internal frame components
of the natural gauge connection for the group of $p$-dimensional
internal frame rotations. It satisfies the properties
\begin{equation}
     \rho_{\lambda\mu\nu} = - \rho_{\lambda\nu\mu}, \quad
     \bot^{\rho} {_{\lambda}} \ \rho_{\rho\mu\nu} = 0 =
     \bot^{\rho} {_{\lambda}} \ \rho_{\mu\rho\nu},
\end{equation}
whereas the internal curvature tensor (1) satisfies the usual
Riemann symmetry properties and the Ricci contractions
\begin{equation}
     R_{\mu\nu} = R_{\mu\sigma\nu} {^{\sigma}}, \quad R =
     R_{\sigma} {^{\sigma}},
\end{equation}
with
\begin{equation}
     \bot^{\sigma} {_{\beta}} \ R_{\sigma\lambda\mu\nu} = 0,
     \quad  \bot^{\sigma} {_{\beta}} \ R_{\sigma\mu} = 0.
\end{equation}
From the fundamental tensor and the Ricci contractions (5) one can
define the internal adjusted Ricci tensor as
\begin{equation}
     \widetilde{R}_{\mu\nu} \equiv R_{\mu\nu} - \frac{1}{2(p -
     1)} \ R \ n_{\mu\nu},
\end{equation}
where $p$ is the dimension of the imbedded $p$-surface. As pointed
out in \cite{5}, for the special case $p = 2$ of a two-dimensional
imbedded surface (that applies to string theory, for which this
work is concerned), the adjusted Ricci tensor (7) vanishes
identically:
\begin{equation}
     \widetilde{R}_{\mu\nu} \equiv R_{\mu\nu} - \frac{1}{2} R \
     n_{\mu\nu} = 0.
\end{equation}
Eq. \ (8) will imply, as we shall see below, that the inner
curvature scalar given in (5) can not give any effective
contribution in a variational principle.

Additionally, the second fundamental tensor is defined by \cite{5}
\begin{equation}
     K_{\lambda\mu}{^{\nu}} = n^{\sigma}_{\mu} \
     \overline{\nabla}_{\lambda} \ n^{\nu}_{\sigma} =
     K_{(\lambda\mu)}{^{\nu}},
\end{equation}
with its property of tangentiality of the first two indices and
orthogonality of the last, and one defines then the curvature
vector as
\begin{equation}
     K^{\rho} = K_{\nu}{^{\nu\rho}}, \qquad n^{\mu}_{\rho} \
     K^{\rho} = 0,
\end{equation}
and
\begin{equation}
     K_{\nu}{^{\rho\nu}} = 0.
\end{equation}
The third fundamental tensor
\begin{equation}
     \Theta_{\kappa\lambda\mu}{^{\nu}} = n^{\rho}_{\lambda} \
     n^{\sigma}_{\mu} \ \bot^{\nu}_{\tau} \
     \overline{\nabla}_{\kappa} \ K_{\rho\sigma}{^{\tau}} =
     \Theta_{\kappa(\lambda\mu)}{^{\nu}},
\end{equation}
satisfies
\begin{eqnarray}
    \overline{\nabla}_{\kappa} \ K_{\lambda\mu}{^{\nu}} \!\! & = &
    \!\! \Theta_{\kappa\lambda\mu}{^{\nu}} + 2 K_{\kappa}
    {^{\sigma}}_{(\lambda} K_{\mu )\sigma}{^{\nu}} -
    K_{\kappa}{^{\nu}}_{\sigma} \ K_{\lambda\mu}{^{\sigma}},
    \nonumber \\
    2 \Theta_{[\kappa\lambda]\mu}{^{\nu}} \!\! & = & \!\!
    n_{\kappa}^{\rho} \ n_{\lambda}^{\sigma} \ n_{\mu}^{\tau}
    \ \bot_{\gamma}{^{\nu}} \ B_{\rho\sigma}{^{\gamma}}_{\tau},
\end{eqnarray}
where $B_{\rho\sigma}{^{\gamma}}_{\tau}$ is the background Riemann
curvature tensor.

Finally, we need to obtain from the Bianchi identity for the
internal curvature (in a general background) \cite{5},
\[
     n_{[\kappa}{^{\nu}} \ n^{\rho}_{\lambda} \ \overline{\nabla}_{\mu
     ]} \ R_{\nu\rho}{^{\sigma\tau}} = 2 R_{[\kappa\lambda}{^{\nu
     [\tau}} K_{\mu ]\nu}{^{\sigma ]}},
\]
its contracted version
\begin{equation}
     \overline{\nabla}_{\mu} (2 R^{\mu\nu} - R n^{\mu\nu}) =
     (2R^{\sigma\rho} - R n^{\sigma\rho}) K_{\sigma\rho}{^{\nu}},
\end{equation}
where the tangentiality and orthogonality of
$K_{\lambda\mu}{^{\nu}}$ and the relation
$\overline{\nabla}_{\lambda} n_{\mu\nu} = 2K_{\lambda(\mu\nu)}$
have been used.

For more details about this section, see directly Refs. \cite{5,6}. \\

\noindent {\uno III. Preliminaries of the deformation dynamics}
\vspace{1em}

Within the covariant scheme given by Carter \cite{6} for the
fluctuation dynamics, it is known that
\begin{equation}
     \delta \sqrt{-\gamma} = \frac{1}{2} \sqrt{-\gamma} \
     n^{\mu\nu} \ \delta g_{\mu\nu},
\end{equation}
where $\gamma$ is the determinant of the embedded surface metric,
and the variation of the background metric is given by its Lie
derivative with respect to the deformation vector field $\xi^{\mu}
= \delta X^{\mu}$ of the embedding,
\begin{equation}
     \delta g_{\mu\nu} = \nabla_{\mu} \ \xi_{\nu} + \nabla_{\nu} \
     \xi_{\mu}.
\end{equation}
Furthermore, in \cite{2} it is shown, taking into account the
gauge nature of $\rho$, that
\begin{eqnarray}
     \delta R_{\kappa\lambda} {^{\mu}}_{\nu} \!\! & = & \!\! 2
     n_{\sigma}{^{\mu}} \ n_{\nu}{^{\tau}} \ n_{[\lambda}{^{\pi}} \
     \overline{\nabla}_{\kappa ]} \ \delta \ \rho_{\pi}
     {^{\sigma}}_{\tau}, \nonumber \\
     \delta  R_{\mu\nu} \!\! & = & \!\! 2 n_{\sigma}{^{\kappa}} \
     n_{\nu}{^{\tau}} \ n_{[\mu}{^{\pi}} \ \overline{\nabla}_{\kappa ]}
     \ \delta \ \rho_{\pi} {^{\sigma}}_{\tau}, \nonumber \\
     n^{\mu\nu} \ \delta  R_{\mu\nu} \!\! & = & \!\!
     \overline{\nabla}_{\mu} \ \psi^{\mu}_{top},
\end{eqnarray}
where
\begin{equation}
     \psi^{\mu}_{top} = n^{\alpha\beta} \ \delta \
     \rho^{\mu}_{\alpha\beta} - n^{\alpha}_{\beta} \ n^{\mu\tau} \
     \delta \ \rho_{\alpha} {^{\beta}}_{\tau}.
\end{equation}
Using Eqs.\ (17), and (18), and considering that $R = n^{\mu\nu}
R_{\mu\nu}$, in \cite{2} it is shown that the variation of a
Lagrangian term proportional to the inner curvature scalar of the
imbedded $p$-surface
\begin{equation}
     \chi = \sigma_{1} \int \sqrt{-\gamma} R \ d \overline{\Sigma},
\end{equation}
is given by
\begin{equation}
     \delta \ \chi = \sigma_{1} \int \sqrt{-\gamma} \ \left(
     \frac{1}{2} R \ n^{\mu\nu} - R^{\mu\nu} \right) \delta
     g_{\mu\nu} \ d \overline{\Sigma} + \sigma_{1} \int
     \overline{\nabla}_{\mu} \ \psi^{\mu}_{top} \ d
     \overline{\Sigma};
\end{equation}
where $\sigma_{1}$ is a fixed parameter.

In general, $\frac{1}{2} R \ n^{\mu\nu} - R^{\mu\nu}$ does no
vanish for a geometry of arbitrary dimension, however, from Eq.\
(8), the adjusted Ricci tensor vanishes identically for string
theory, and $\chi$ does not give dynamics to such objects. Hence
$\chi$ is a topological invariant for string theory (for
appropriate boundary conditions), that geometrically corresponds
to the number of handles of the world-surface, the so called Euler
characteristic.

Considering the orthogonal gauge $n^{\mu}_{\nu} \xi^{\nu} = 0$,
which removes the nonphysically observable tangential of the
deformation \cite{6} (and that we consider throughout this work),
the first term in Eq.\ (20) can be rewritten (using Eq.\ (16)) as
\begin{eqnarray}
     \big( \frac{1}{2} R \ n^{\mu\nu} - R^{\mu\nu} \big) \delta
     g_{\mu\nu} \!\! & = & \!\! \xi^{\nu} \ \overline{\nabla}_{\mu}
     (2 R^{\mu}{_{\nu}} - R \ n^{\mu}{_{\nu}}) \nonumber \\
     \!\! & = & \!\! (2 R^{\sigma\rho} - R n^{\sigma\rho})
     K_{\sigma\rho\nu} \xi^{\nu},
\end{eqnarray}
where the last equality follows from the contracted Bianchy
identity (14). Thus, Eq.\ (21) implies from Eq.\ (20) that the
equation of the motion for $\chi$ is given by
\begin{equation}
\sigma_{1} (2 R^{\sigma\rho} - R n^{\sigma\rho})
     K_{\sigma\rho\nu} = 0.
\end{equation}
\\

\noindent {\uno IV. Fluctuations of a topological invariant in an
arbitrary background}
\vspace{1em}

The idea now is to calculate the deformations of the equations of
motion (22) for an arbitrary $p$-brane and to impose on the
deformation equations thus obtained the condition (8) that
identifies the two-dimensional world-surface. The result will be
that the equations governing the fluctuations of the topological
invariant admit solutions different to the trivial ones, in spite
of the null dynamics at the level of the unperturbed equations of
motion.

Hence, the variations of the equations (22) are given by
\begin{equation}
     \sigma_{1} K_{\sigma\rho\nu} \delta (2R^{\sigma\rho} - R
     n^{\sigma\rho}) + \sigma_{1} (2R^{\sigma\rho} - R
     n^{\sigma\rho}) \delta K_{\sigma\rho\nu} = 0,
\end{equation}
where the second term will be not worked out, because will vanish
under the condition (8). On the other hand, the variation of the
adjusted Ricci tensor in the first term in Eq.\ (23) can be
written in terms of the expressions (16), and (17) as
\begin{eqnarray}
     \delta (2R^{\sigma\rho} - R \ n^{\sigma\rho}) \!\! & = & \!\! 2
     \ \overline{\cal C}^{\mu\nu\sigma\rho} \delta R_{\mu\nu} +
     (R^{\alpha\beta} n^{\sigma\rho} - 2 n^{\rho\alpha}
     R^{\sigma\beta}) \delta g_{\alpha\beta} \nonumber \\
     \!\! &  & \!\! - (2 R^{\beta\rho} - R n^{\beta\rho})
     n^{\sigma\alpha} \delta g_{\alpha\beta},
\end{eqnarray}
where the third term vanishes under the condition (8), and the
other ones are reduced to
\begin{equation}
     \delta (2R^{\sigma\rho} - R \ n^{\sigma\rho}) =
     \overline{\cal C}^{\mu\nu\sigma\rho} (2\delta  R_{\mu\nu} -
     R \ \delta g_{\mu\nu}),
\end{equation}
under the same condition (8). $\overline{\cal
C}^{\mu\nu\sigma\rho}$ is the called hyper-Cauchy tensor
introduced by Carter \cite{7}:
\begin{eqnarray}
     \overline{\cal C}^{\mu\nu\sigma\rho} \!\! & \equiv & \!\!
     n^{\sigma(\mu} n^{\nu)\rho} -\frac{1}{2} n^{\mu\nu} \
     n^{\sigma\rho} \\
     \!\! & =  & \!\! \overline{\cal C}^{(\mu\nu)(\sigma\rho)}.
     \nonumber
\end{eqnarray}
From Eq.\ (25) we immediately conclude that, although the adjusted
Ricci tensor does vanish for string theory, its fluctuations do
not necessarily. In fact, the existence of nontrivial fluctuations
for a topological term is a generic issue in all physical systems
\cite{2}, and one may make the corresponding for any system of
interest.

For our proposes, we need to work out more Eqs.\ (23), in order to
express them explicitly in terms of the embedding deformation
vector $\xi^{\mu}$, and thus it is essential to know the
deformations of the frame vectors $\{ i^{\mu}_{A} \}$ tangential
to the world-surface:
\begin{equation}
     \delta \ i^{\mu}_{A} = i^{\alpha}_{A} \ K_{\alpha}
     {^{\mu}}_{\lambda} \ \xi^{\lambda},
\end{equation}
which implies that the deformations of the internal connection
$\rho_{\lambda} {^{\mu}}_{\nu}$ is given, after a long
calculation, by
\begin{eqnarray}
     \delta \rho_{\lambda} {^{\mu}}_{\nu} \!\! & = & \!\!
     K_{\lambda\nu\alpha} \overline{\nabla}^{\mu} \ \xi^{\alpha} -
     K_{\lambda} {^{\mu}}_{\alpha} \overline{\nabla}_{\nu} \
     \xi^{\alpha} - n^{\mu}_{\rho} \ n^{\sigma}_{\nu} \
     n^{\tau}_{\lambda} \ {\cal B}^{\rho}{_{\sigma\tau\alpha}} \
     \xi^{\alpha} \\
     \!\! & = & \!\! \overline{\nabla}^{\mu} (K_{\nu\lambda\alpha} \
     \xi^{\alpha}) - \overline{\nabla}_{\nu} (K^{\mu} {_{\lambda\alpha}} \
     \xi^{\alpha}) + 2 K_{\nu} {^{\rho}}_{(\lambda} K^{\mu )}
     {_{\rho\alpha}} \ \xi^{\alpha} - 2 K^{\mu\rho} {_{(\lambda}} K_{\nu
     )\rho\alpha} \ \xi^{\alpha}. \nonumber
\end{eqnarray}
Note that Eq.\ (28) is a manifestly tensorial expression for the
deformation of the connection $\rho$, with support confined to the
worldsheet; $\delta \rho$ also turns out to be antisymmetric on
the indices $\mu,\nu$, such as $\rho_{\lambda}{^{\mu}}_{\nu}$
itself. The last equality in Eq.\ (28) is obtained using the
relation (13), and it shows in a manifest form that the
deformation $\delta \rho$ is expressed in terms of the fundamental
field of the deformation dynamics, $K_{\alpha}{^{\mu}}_{\lambda}
\xi^{\lambda}$, which is clear also in Eq.\ (27), and will be a
rule throughout the present treatment.

Therefore, from Eqs.\ (17), and (28), we can determine the
(tangential projection) of the deformation of the internal Ricci
tensor,
\begin{eqnarray}
     n^{\mu}_{\alpha} \ n^{\nu}_{\beta} \ \delta R_{\mu\nu} \!\!
     & = & \!\! n^{\mu}_{\alpha} \ n^{\nu}_{\beta} [
     \overline{\nabla}_{\sigma} \overline{\nabla}^{\sigma} (K_{\mu\nu\lambda} \
     \xi^{\lambda}) - \overline{\nabla}_{\sigma}
     \overline{\nabla}_{\nu} (K_{\mu}{^{\sigma}}_{\lambda} \
     \xi^{\lambda}) - \overline{\nabla}_{\mu} \overline{\nabla}^{\sigma}
     (K_{\nu\sigma\lambda} \  \xi^{\lambda}) \nonumber \\
     \!\! & & \!\! + \overline{\nabla}_{\mu}
     \overline{\nabla}_{\nu} (K_{\alpha} \ \xi^{\alpha}) + (2
     K_{\nu\sigma\rho} K^{\sigma\tau\rho} - K_{\sigma}
     K_{\nu}{^{\tau\sigma}}) K_{\mu\tau\lambda} \ \xi^{\lambda}
     \nonumber \\
     \!\! & & \!\! - (K_{\mu\nu\rho} K^{\sigma\tau\rho} + K_{\nu}
     {^{\sigma}}_{\rho} K_{\mu}{^{\tau\rho}}) K_{\sigma\tau\lambda}
     \ \xi^{\lambda}].
\end{eqnarray}
In this manner, Eqs.\ (25), and (29), convert Eq.\ (23) into a
second-order linear differential equations for the deformation
vector $(K_{\mu\nu\alpha}) \xi^{\alpha}$, that we can express in a
compact form as
\begin{equation}
     ({\cal O} \ \xi^{\alpha})_{\nu} = 0,
\end{equation}
where the linear operator $\cal O$ maps vector fields into
themselves.

Let us prove now the self-adjointness of the linear operator $\cal
O$ in Eq.\ (30) \cite{8}, and we consider two vector fields
$\acute{\xi}^{\alpha}$, and $\xi^{\alpha}$,which will be
identified finally with solutions of Eqs.\ (30).

Considering Eq.\ (25), the contraction of Eq.\ (23) or (30) with
$\acute{\xi}^{\nu}$ gives a term that does not involve
differential operators:
\begin{equation}
     (\acute{\xi}^{\nu} K_{\sigma\rho\nu}) R \ \overline{\cal
     C}^{\mu\nu\sigma\rho} \delta g_{\mu\nu} = -2 R \ \overline{\cal
     C}^{(\mu\nu)(\sigma\rho)} (K_{\sigma\rho\beta} \
     \acute{\xi}^{\beta}) (K_{\mu\nu\alpha} \xi^{\alpha}),
\end{equation}
where we have employed the orthogonal gauge. For the terms
involving differential operators we can use differential
identities of the form
\begin{eqnarray}
     K_{\sigma\rho\beta} \ \acute{\xi}^{\beta} \ \overline{\nabla}_{\lambda}
     \overline{\nabla}^{\lambda} (K_{\mu\nu\alpha} \xi^{\alpha})
     \!\! & \equiv & \!\! \overline{\nabla}_{\lambda} [
     K_{\sigma\rho\beta} \ \acute{\xi}^{\beta} \ \overline{\nabla}^{\lambda}
     (K_{\mu\nu\alpha} \ \xi^{\alpha}) -
     \overline{\nabla}^{\lambda} (K_{\sigma\rho\beta} \
     \acute{\xi}^{\beta}) K_{\mu\nu\alpha} \ \xi^{\alpha} ]
     \nonumber \\
     \!\! & & \!\! + \overline{\nabla}_{\lambda}
     \overline{\nabla}^{\lambda} (K_{\sigma\rho\beta} \
     \acute{\xi}^{\beta}) K_{\mu\nu\alpha} \ \xi^{\alpha},
\end{eqnarray}
and hence, considering Eqs.\ (30) and (31), we can find, after
some arrangements, that
\begin{equation}
     \acute{\xi}^{\nu} ({\cal O} \ \xi^{\alpha})_{\nu} - ({\cal O} \
     \acute{\xi}^{\nu})_{\alpha} \xi^{\alpha} =
     \overline{\nabla}_{\mu} \ \overline{J}^{\mu},
\end{equation}
or more explicitly
\begin{equation}
     \sigma_{1} \ K_{\sigma\rho\nu} \ \acute{\xi}^{\nu} \ \overline{\cal
     C}^{\mu\nu\sigma\rho} (2\delta R_{\mu\nu} - R \delta g_{\mu\nu})
     - \sigma_{1} \ \overline{\cal C}^{\mu\nu\sigma\rho}
     (2\delta R'_{\mu\nu} - R \delta g'_{\mu\nu})
     K_{\sigma\rho\alpha} \ \xi^{\alpha} = \overline{\nabla}_{\mu} \
     \overline{J}^{\mu},
\end{equation}
where $\delta R'_{\mu\nu}$ corresponds to $\delta R_{\mu\nu}$ with
argument $\acute{\xi}^{\alpha}$, and similarly for $\delta
g'_{\mu\nu}$. Moreover,
\begin{eqnarray}
     \frac{1}{2} \overline{J}^{\mu} \!\! & = & \!\!
     K^{\lambda\tau}{_{\beta}} \ \acute{\xi}^{\beta} \
     \overline{\nabla}^{\mu} (K_{\lambda\tau\alpha} \
     \xi^{\alpha}) - K^{\nu\lambda}{_{\beta}} \ \acute{\xi}^{\beta} \
     \overline{\nabla}_{\lambda} (K_{\nu}{^{\mu}}_{\alpha} \
     \xi^{\alpha}) \nonumber \\
     \!\! & & \!\! - K^{\mu\nu}{_{\beta}} \ \acute{\xi}^{\beta} \
     \overline{\nabla}_{\lambda} (K_{\nu}{^{\lambda}}_{\alpha} \
     \xi^{\alpha}) + K^{\mu\nu}{_{\beta}} \ \acute{\xi}^{\beta} \
     \overline{\nabla}_{\nu} (K_{\alpha} \ \xi^{\alpha}) \nonumber
     \\
     \!\! & & \!\! - \overline{\nabla}_{\nu} (K^{\mu\nu}{_{\beta}} \
     \acute{\xi}^{\beta}) K_{\alpha} \ \xi^{\alpha} +
     K^{\lambda\nu\mu} K_{\lambda\nu\beta} K_{\alpha} (\acute{\xi}^{\beta}
     \xi^{\alpha}) \nonumber \\
     \!\! & & \!\! - K_{\beta} \ \acute{\xi}^{\beta} \
     \overline{\nabla}^{\mu} (K_{\alpha} \ \xi^{\alpha}) -
     (\acute{\xi}^{\alpha} \leftrightarrow \xi^{\alpha});
\end{eqnarray}
in this manner, $\overline{J}^{\mu}$ contains all arguments of
total divergences such as the first term on the right-hand side of
Eq.\ (32).

Finally, from Eqs.\ (33), or (34), we conclude that, if
$\acute{\xi}^{\alpha}$ and $\xi^{\alpha}$ correspond to a pair of
solutions admitted by the deformation dynamics (30),
$\overline{J}^{\mu}$ is worldsheet covariantly conserved
\begin{equation}
      \overline{\nabla}_{\mu} \ \overline{J}^{\mu} = 0.
\end{equation}
As we shall see, $\overline{J}^{\mu}$ will correspond to the
integral kernel of a symplectic structure for the topological
term, and the property (36) will make sense then. To finish this
section, it is worth to point out that $\overline{J}^{\mu}$ and
its property (36) emerge directly from the nontrivial fluctuations
of the topological term, which will constitute an important
ingredient in the final setting of our results. \\

\noindent {\uno V. The symplectic structure for $\chi$}
\vspace{1em}

Once we know the deformations of the internal conecction (28), one
can find explicitly the deformation of the symplectic potential
(18), which constitute the integral kernel of the symplectic
structure of the topological term \cite{2}. Considering Eq.\ (18),
we have
\[
     \delta \psi^{\mu}_{top} = \delta \ n^{\alpha\beta} \ \delta \
     \rho_{\alpha}{^{\mu}}_{\beta} - n^{\alpha}_{\beta} \ \delta \
     n^{\mu\tau} \ \delta \ \rho_{\alpha}{^{\beta}}_{\tau} -
     n^{\mu\tau} \ \delta \ n^{\alpha}_{\beta} \ \delta \
     \rho_{\alpha}{^{\beta}}_{\tau}, \nonumber
\]
where the last term vanishes because $\delta n^{\alpha}_{\beta}$
is proportional to $\bot^{\sigma}_{\beta}$ (orthogonal to the
world-surface), and $\delta \rho_{\alpha}{^{\beta}}_{\tau}$ is
proportional to $n^{\beta}_{\lambda}$ (tangential to the
world-surface). Using now Eqs.\ (15), and (28), one obtains that
\begin{eqnarray}
     \delta (\sqrt{-\gamma} \ \psi^{\mu}_{top}) \!\! & = & \!\! 2
     \sqrt{-\gamma} [ K^{\lambda\tau}{_{\beta}} {\xi}^{\beta}
     \ \overline{\nabla}^{\mu} (K_{\lambda\tau\alpha}
     \xi^{\alpha}) - K^{\nu\lambda}{_{\beta}}  {\xi}^{\beta} \
     \overline{\nabla}_{\lambda} (K_{\nu}{^{\mu}}_{\alpha}
     \xi^{\alpha}) \nonumber \\
     \!\! & & \!\! - K^{\mu\nu}{_{\beta}} {\xi}^{\beta} \
     \overline{\nabla}_{\lambda} (K_{\nu}{^{\lambda}}_{\alpha} \
     \xi^{\alpha}) + K^{\mu\nu}{_{\beta}} {\xi}^{\beta} \
     \overline{\nabla}_{\nu} (K_{\alpha}  \xi^{\alpha}) \nonumber
     \\
     \!\! & & \!\! - \overline{\nabla}_{\nu} (K^{\mu\nu}{_{\beta}} \
     {\xi}^{\beta}) K_{\alpha}  \xi^{\alpha} +
     K^{\lambda\nu\mu} K_{\lambda\nu\beta} K_{\alpha} ({\xi}^{\beta}
     \xi^{\alpha}) \nonumber \\
     \!\! & & \!\! - K_{\beta} {\xi}^{\beta} \
     \overline{\nabla}^{\mu} (K_{\alpha}  \xi^{\alpha})];
\end{eqnarray}
which is essentially the current found in Eq.\ (35) if we set up
$\acute{\xi}^{\alpha} = \xi^{\alpha}$ \cite{8}. Hence, the
symplectic structure  $\omega$ for $\chi$ can be written as
\cite{2}
\begin{equation}
     \omega = \int_{\Sigma} \delta (\sqrt{-\gamma} \ \psi^{\mu}_{top})
     d \Sigma_{\mu} = \int_{\Sigma} \sqrt{-\gamma} \
     \overline{J}^{\mu} \ d \Sigma_{\mu},
\end{equation}
the first form proves, as we already know \cite{2}, that $\omega$
is closed, and the second one proves that is independent on the
choice of $\Sigma$, due to the property (36).

In \cite{1,2,8}, $\Sigma$ is defined as {\it `` a spacelike
section of the worldsheet corresponding to a Cauchy surface for
the configuration of the string."} What is $\Sigma$ for a
topological term without dynamics? The nontrivial deformation
dynamics comes to rescue: $\omega$ is finally defined in terms of
field deformations (which, in turn, are defined in terms of the
solutions for the deformation dynamics); therefore, $\Sigma$ will
be in this case a spacelike surface for the configuration of the
deformation fields appearing in $\omega$. \\

\noindent {\uno VI. $\omega$ and the symplectic structure of an
Abelian gauge theory} \vspace{1em}

In this section we shall show the analogy between the symplectic
structure found in the present treatment for the Euler
characteristic, and that found in \cite{9} for an Abelian gauge
theory within a covariant canonical formalism.

In \cite{9}, it is found that
\begin{equation}
     \hat{\omega} = \int_{\Sigma} {\rm Tr} (\delta \ F^{\mu\alpha} \ \delta \
     A_{\mu}) d\Sigma_{\alpha},
\end{equation}
is a covariant and gauge invariant symplectic structure for
Yang-Mills theory; $F_{\mu\nu}$ is the usual Yang-Mills curvature,
$A_{\mu}$ the corresponding gauge connection, and $\Sigma$ a
spacelike hypersurface. As we know, $F$ is the {\it exterior
derivative} of $A$,
\begin{equation}
     F = {\cal D} A,
\end{equation}
and $F$ is also the solution of the equations
\begin{equation}
     [\nabla_{\mu}, F^{\mu\nu}] = 0.
\end{equation}
Eqs.\ (39), (40), and (41) contain, of course, the Abelian gauge
theory as a particular case.

On the other hand, within the differential geometry of the
imbedded surface that we are employing, the fact that the
embedding has dimension 2 (and that applies for the present case),
implies that the inner rotation group is Abelian \cite{5}, the
analogue of the Abelian (internal) gauge group. Moreover, in this
particular case the imbedding two-surface is characterized by the
antisymmetric unit tangent element tensor given by \cite{5}
\begin{equation}
     {\cal E}^{\mu\nu} = {\cal E}^{[\mu\nu]} = {\cal E}^{AB} \
     i^{\mu}_{A} \ i^{\nu}_{B},
\end{equation}
${\cal E}^{AB}$ being the constant components of the standard
two-dimensional flat space alternating tensor; ${\cal E}^{\mu\nu}$
will be, of course, the analogue of $F^{\mu\nu}$. However, the
analogy is not only at the level of the antisymmetry property of
$\cal E$ and $F$, since using the tangential derivative of $\cal
E$ given in \cite{5} by
\begin{equation}
     \overline{\nabla}_{\sigma} {\cal E}^{\mu\nu} = 2
     K_{\sigma\tau}{^{[\nu}} {\cal E}^{\mu ]\tau},
\end{equation}
we can obtain the ``wold-surface divergence" of $\cal E$
contracting Eq.\ (43) as
\begin{equation}
     \overline{\nabla}_{\mu} {\cal E}^{\mu\nu} =
     K_{\mu\tau}{^{\nu}} \ {\cal E}^{\mu\tau} - K_{\mu\tau}{^{\mu}}
     \ \rho_{\mu}{^{\alpha}}_{\beta} \ {\cal E}^{\beta}{_{\alpha}}
     = 0,
\end{equation}
the first term vanishes because of the symmetry of $K$ in
$(\mu\tau)$ and the antisymmetry of ${\cal E}^{\mu\tau}$; the
second term because $K_{\mu\tau}{^{\mu}} = 0$ (see Eq.\ (11)).
Equation (44) is then the analogue of the field equations (41) for
the Abelian case.

Continuing, in \cite{5} it is shown that $R{\cal E}$ is the {\it
exterior derivative} of the (locally defined, frame gauge
dependent) {\it one-form} $\rho_{\mu} =
\rho_{\mu}{^{\alpha}}_{\beta} \ {\cal E}^{\beta}{_{\alpha}}$ in
the embedding surface,
\begin{equation}
     R \ {\cal E} = \overline{\partial} \rho,
\end{equation}
where $\overline{\partial}$ is the exterior derivative projected
on the imbedding surface \cite{5}; in components, Eq.\ (45) takes
the form \cite{5}
\begin{equation}
     R \ {\cal E}_{\kappa\lambda} = 2 n_{[\lambda}{^{\sigma}}
     \overline{\nabla}_{\kappa ]} \ \rho_{\sigma},
\end{equation}
which is the analogue of (40), and $\rho_{\mu}$ is thus the
analogue of the gauge connection $A_{\mu}$.

In order to complete our pretended analogy, it remains to be seen
if our $\omega$ takes the form (39) given for a gauge theory. This
is effectively the case: the topological invariant can be
rewritten as \cite{5,6}
\begin{equation}
     \chi = \int R \ d \overline{\Sigma} = \int
     \overline{\nabla}_{\mu} ({\cal E}^{\mu\nu} \ \rho_{\nu}) d
     \overline{\Sigma},
\end{equation}
and in \cite{2}, considering again the gauge nature of $\rho$, we
shown that the variation of $\chi$ reads
\begin{equation}
     \delta \ \chi = \int \overline{\nabla}_{\mu}
     ({\cal E}^{\mu\nu} \ \delta \ \rho_{\nu}) d
     \overline{\Sigma},
\end{equation}
and hence, we identified $(\sqrt{-\gamma}) {\cal E}^{\mu\nu}
\delta \rho_{\nu}$ as a symplectic potential for $\chi$, which
corresponds exactly to $\psi^{\mu}_{top}$ described above
\cite{2}. This implies that $\omega$ can be written also as
\begin{eqnarray}
     \omega = \int_{\Sigma} \delta (\psi^{\mu}_{top}) d
     \overline{\Sigma}_{\mu} \!\! & = & \!\! \int_{\Sigma} \delta
     (\sqrt{-\gamma} \ {\cal E}^{\mu\nu} \ \delta \ \rho_{\nu}) d
     \overline{\Sigma}_{\mu} \nonumber \\
     \!\! & = & \!\!  \int_{\Sigma} \delta (\sqrt{-\gamma} \
     {\cal E}^{\mu\nu}) \ \delta \ \rho_{\nu} \ d
     \overline{\Sigma}_{\mu},
\end{eqnarray}
where we have considered in the last equality the nilpotency of
the exterior derivative \cite{2}. Therefore $\omega$ takes exactly
the required form (39).

Therefore $\omega$ mimics entirely $\hat{\omega}$ in the relevant
aspects of its mathematical structure. \\

\noindent {\uno VII. $\omega$ mimics $\hat\omega$ in the symmetry
properties} \vspace{1em}

Following the idea of the covariant canonical formalism of
preserving manifestly the relevant symmetries of the theory, in
\cite{9} it is proved that the symplectic form (39) is a covariant
and gauge invariant geometrical structure for the gauge theory.
Specifically, under the ordinary gauge transformation,
\begin{equation}
     A_{\mu} \rightarrow A_{\mu} + \partial_{\mu} \phi,
\end{equation}
$\delta A_{\mu}$ and $\delta F^{\mu\nu}$ transform homogeneously,
and thus $\hat{\omega}$ in Eq.\ (39) is gauge invariant. Moreover,
the gauge directions on the phase space are defined by the
transformation
\begin{equation}
     \delta A_{\mu} \rightarrow \delta A_{\mu} + \partial_{\mu}
     \phi,
\end{equation}
and $\hat{\omega}$ proves to be also invariant under (51), which
defines $\hat{\omega}$ on the reduced phase space $Z \equiv
\widehat{Z}/G$, where $\widehat{Z}$ is the space of solutions of
Eqs.\ (41), and $G$ the group of gauge transformations \cite{9}.

We analyze now the analogous situation in our present case for the
topological invariant $\chi$. Since the analogy seems to be filled
out, we can guess the corresponding gauge transformation for the
gauge connection $\rho_{\nu}$ (in an arbitrary curved background),
\begin{equation}
     \rho_{\nu} \rightarrow \rho_{\nu} + \overline{\nabla}_{\nu}
     \phi,
\end{equation}
where $\phi$ is an arbitrary scalar field;
$\overline{\nabla}_{\nu}$ is the appropriate derivative since
$\rho_{\nu}$ is of support confined to the imbedding surface (see
Eq.\ (3)). Considering the expression (47) for $\chi$, one easily
verifies that effectively (52) is a symmetry of the original
action:
\begin{equation}
     \chi' = \int \overline{\nabla}_{\mu} [ {\cal E}^{\mu\nu}
     (\rho_{\nu} + \overline{\nabla}_{\nu}\phi)] d
     \overline{\Sigma} = \chi + \int {\cal E}^{\mu\nu}
     \overline{\nabla}_{\mu} \ \overline{\nabla}_{\nu} \ \phi \ d
     \overline{\Sigma},
\end{equation}
where we have considered Eq.\ (44). Now, taking into account that
$\nabla_{[\mu}\nabla_{\nu ]} \phi = 0$, we can find easily that
\begin{equation}
     \overline{\nabla}_{[\mu} \overline{\nabla}_{\nu ]} \phi =
     K_{[\mu}{^{\sigma}}_{\nu ]} \ \overline{\nabla}_{\sigma}
     \phi,
\end{equation}
and considering that additionally ${\cal E}^{\mu\nu}
K_{\alpha\beta\nu} = 0$, the integrand in the second term in (53)
vanishes, and then $\chi' = \chi$. Note that $\chi$ is strictly
invariant, it does not change by a total divergence.

Additionally, $R{\cal E}_{\mu\nu}$ in Eq.\ (46) is also invariant,
as expected, under the gauge transformation (52),
\begin{equation}
     (R{\cal E}_{\kappa\lambda})' = R \ {\cal E}_{\kappa\lambda} +
     2 [ \overline{\nabla}_{[\kappa} \overline{\nabla}_{\lambda ]}
     \phi + K_{[\lambda}{^{\sigma}}_{\kappa ]}
     \overline{\nabla}_{\sigma} \phi ],
\end{equation}
where the second term vanishes due again to (54).

We show now that $\omega$ in (49) retains these symmetries. Since
$\delta \rho_{\nu}$ transforms homogeneously under (52), $\omega$
is automatically gauge invariant ($\delta(\sqrt{-\gamma} {\cal
E}^{\mu\nu})$ does not depend on $\rho_{\nu}$). The situation
becomes interesting when we consider the gauge transformation in
{\it field space},
\begin{equation}
     \delta \rho_{\nu} \rightarrow \delta \rho_{\nu} +
     \overline{\nabla}_{\nu} \phi,
\end{equation}
the analogue of (51). $\omega$ undergoes the transformation
\begin{eqnarray}
     \omega' \!\! & = & \!\! \int_{\Sigma} \delta (\sqrt{-\gamma} \
     {\cal E}^{\mu\nu}) (\ \delta \rho_{\nu} + \overline{\nabla}_{\nu}
     \phi) \ d \overline{\Sigma}_{\mu} = \omega + \int_{\Sigma} \delta
     (\sqrt{-\gamma} \ {\cal E}^{\mu\nu}) \ \overline{\nabla}_{\nu} \phi
     \ d \overline{\Sigma}_{\mu} \nonumber \\
     \!\! & = & \!\! \omega + \int_{\Sigma}
     \overline{\nabla}_{\nu} [\phi \ \delta (\sqrt{-\gamma} \
     {\cal E}^{\mu\nu})] \ d \overline{\Sigma}_{\mu} - \int_{\Sigma}
     \phi \overline{\nabla}_{\nu} [\delta (\sqrt{-\gamma} \
     {\cal E}^{\mu\nu})] \ d \overline{\Sigma}_{\mu},
\end{eqnarray}
therefore, this equation implies that $\omega$ will change by a
total divergence, the second term on the right hand-side, if the
last one vanishes. Since $\phi$ is an arbitrary field, the only
possibility is that $\overline{\nabla}_{\nu} (\delta
(\sqrt{-\gamma} \ {\cal E}^{\mu\nu}))$ vanishes. Let us prove that
this is effectively the situation.

We take first the variation of Eq.\ (44):
\begin{equation}
     \delta (\overline{\nabla}_{\mu} {\cal E}^{\mu\nu}) =
     \overline{\nabla}_{\mu} \ \delta \ {\cal E}^{\mu\nu} + {\cal
     E}^{\lambda\nu} \ n^{\alpha}_{\mu} \ \delta \
     \Gamma^{\mu}_{\alpha\lambda} + {\cal E}^{\lambda\nu} \
     K_{\lambda}{^{\rho\sigma}} \ \delta \ g_{\rho\sigma} = 0,
\end{equation}
where we have considered Eq.\ (43), the antisymmetry of ${\cal
E}^{\mu\nu}$, and the symmetry of $\delta
\Gamma^{\nu}_{\alpha\lambda}$ in $(\alpha\lambda)$. On the other
hand, considering that
\begin{equation}
     \delta \Gamma^{\mu}_{\alpha\lambda} = \frac{1}{2} g^{\mu\rho}
     (\nabla_{\alpha} \delta \ g_{\lambda\rho} + \nabla_{\lambda}
     \delta \ g_{\alpha\rho} - \nabla_{\rho} \delta \
     g_{\alpha\lambda}),
\end{equation}
one can show that
\begin{equation}
     n^{\beta}_{\lambda} n^{\alpha}_{\mu} \ \delta
     \Gamma^{\mu}_{\alpha\beta} = \frac{1}{2} n^{\beta}_{\lambda}
     n^{\alpha\rho} \nabla_{\beta} \ \delta \ g_{\alpha\rho} =
     \overline{\nabla}_{\lambda}
     \big( \frac{\delta\sqrt{-\gamma}}{\sqrt{-\gamma}} \big) -
     K_{\lambda}{^{\sigma\rho}} \ \delta \ g_{\sigma\rho},
\end{equation}
where we take into account Eq.\ (15). Thus, from Eqs.\ (58), and
(60), we have
\begin{equation}
     \overline{\nabla}_{\mu} (\delta{\cal E}^{\mu\nu}) = - \frac{{\cal
     E}^{\mu\nu}}{\sqrt{-\gamma}} \overline{\nabla}_{\mu} (\delta
     \sqrt{-\gamma}),
\end{equation}
which implies finally that
\begin{equation}
     \overline{\nabla}_{\mu} [\delta (\sqrt{-\gamma} \
     {\cal E}^{\mu\nu})] = 0,
\end{equation}
as required for the vanishing of the last term on the right
hand-side in Eq.\ (57); the second term  in Eq.\ (57) reduces to
\begin{equation}
     \int_{\Sigma} \overline{\nabla}_{\nu} [\phi \ \delta
     (\sqrt{-\gamma} \ {\cal E}^{\mu\nu})] \ d
     \overline{\Sigma}_{\mu} = \int_{\partial\Sigma}
     \phi \ \delta (\sqrt{-\gamma} \ {\cal E}^{\mu\nu}) \
     d \overline{\Sigma}_{\mu\nu},
\end{equation}
which vanishes if we assume that the field variations are of
compact support at the boundary $\partial\Sigma$ (analogous
boundary conditions are imposed on the field variations in the
case of gauge theory \cite{9}). Under these conditions, $\omega$
has not components along the gauge orbits, and then we have
defined $\omega$ on certain {\it reduced phase space}. Note that,
up to here, we have not defined strictly the phase space, because
we have not equations of motion for $\chi$! On what space the
pre-symplectic form (49), and the subsequent symplectic
structure are defined then?  \\

\noindent {\uno VIII. Phase space and reduced phase space for a
topological invariant}
\vspace{1em}

In the original work on a covariant description of the canonical
formalism \cite{9}, the phase space is defined as {\it the space
of solutions of the equations of motion}, and the reduced  or
physical phase space as the phase space modulo gauge
transformations. However, in the present case, one has only a
topological term without dynamics, and we can not apply such a
definition.

In order to find a covariant definition of phase space for this
case, we can use some ideas known in the literature, and the
analogy here established. We can define first the {\it kinematic
phase space} $Z$ as the space of all smooth $\rho$ connections and
$\cal E$ fields, on which the pre-symplectic form (49) will be
defined. If the equations (44), and (46) for $\cal E$ and $\rho$
are satisfied, we obtain a sub-manifold of $Z$ that we can call
phase space $\overline{Z}$. The reduced phase space $\widehat{Z}$
is obtained then from $\overline{Z}$ by dividing it by the volume
of the symmetry group, we mean $\widehat{Z}$ is constituted by the
set of equivalence classes, where the equivalence relation is
given by the gauge transformations (56): two points on the phase
space $\overline{Z}$ are equivalent if they differ by a gauge
transformation. In conclusion, in a covariant canonical formalism
for the topological invariant $\chi$, the equations (44), and
(46), which characterize the two-dimensional embedding surface,
play the role, in some sense, of the nonexistent equations of
motion. The analogy is in this sense also filled out, since the
analogue of Eqs.\ (44), and (46) for $\chi$ correspond to the
equations of motion
(41) and the Bianchi identify (40) for gauge theory. \\

\noindent {\uno IX. Remarks and prospects}
\vspace{1em}

We remark first that the results established in this work are
valid for a two-dimensional surface embedded in a curved
background of arbitrary dimension. Therefore, considering that the
symplectic structure constructed for $\chi$ will govern finally
the transition between the classical and quantum domains, it is
inevitable to wonder about questions such as the critical
dimension of the background, preservation of the classical
symmetries throughout the quantization process, etc, which turn
out to be of particular interest in string theory. For example, a
way in which the phenomenon of the critical dimension is
manifested in (bosonic) string theory is that the corresponding
Poincar\'e algebra at a quantum level is closed only if the
background dimension is 26. One can ask if $\chi$ has a relevant
effect on this particular question. In this sense, one can study
the unexpected nontrivial realizations of the Poincar\'e algebra
(at classical level) for $\chi$ from its symplectic structure
$\omega$, as a first step in such a direction. In fact, this will
be the subject of forthcoming works.

Considering the point of view, initiated by Dirac, that the
classical $\leftrightarrow$ quantum correspondence of the physical
systems should be formulated in terms of analogies between their
mathematical structures, the similarly established here with an
Abelian gauge theory (perhaps the most studied gauge theory),
allows to reveal the role of the topological terms in a quantum
domain. Conversely, the possible analogy with a more general
non-Abelian gauge theory may be useful in order to construct
topological invariants in higher dimensional systems.

As discussed in \cite{1,2}, there exists another topological
invariant associated with a two-dimensional surface (embedded in a
four-dimensional background), the so called first Chern number of
the normal bundle of the surface, and related geometrically with
the number of self-intersections of the surface. The corresponding
symplectic structure proves to have also the form of an Abelian
gauge theory, in an entirely similar way to the case developed
here; however, explicit calculations will be performed in its
opportunity.

Additionally we comment that in Ref. \cite{10}, it is proved the
conformal invariance of the phase space formulation presented in
this work.

Finally, the results obtained here, can be also achieved using the
symplectic scheme developed by Carter in \cite{7}. \\

\begin{center}
{\uno ACKNOWLEDGMENTS}
\end{center}
\vspace{1em}
 Warm  thanks to M. Montesinos and G. F. Torres del
Castillo for their criticism and comments. The author acknowledges
the support
from the Sistema Nacional de Investigadores (M\'exico). \\


\begin{thebibliography}{}

\setlength{\itemsep}{-.50em}
\bibitem{1} R. Cartas-Fuentevilla, J.\ Math.\ Phys., {\bf 45}, 602
(2004), Preprint math-ph/0404004.
\bibitem{2} R. Cartas-Fuentevilla, and A. Escalante, Trends\ in \ Math.\ Phys.,
Nova publishing, to be published (2004), Preprint math-ph/0404001.
\bibitem{3} M. Mondragon, and M. Montesinos, {\it Covariant canonical formalism for 4-dimensional BF
theory}, preprint (2004).
\bibitem{4} M. Kaku, {\it Strings, conformal fields, and
M-theory}, Springer, New York (1999), Chapter 12.
\bibitem{5} B. Carter, J.\ Geom.\ Phys., {\bf 8}, 53 (1992).
\bibitem{6} B. Carter, 1997 {\it Brane dynamics for treatment of cosmic
strings and vortons}, in {\it Recent Developments in Gravitation
and Mathematics, Proc. 2nd Mexican School on Gravitation and
Mathematical Physics (Tlaxcala, 1996)}
(http://kaluza.physik.uni-konstanz.de/2MS) ed. A. Garcia, C.
Lammerzahl, A. Macias and D. Nu\~{n}ez (Konstanz: Science
Network); Phys.\ Rev.\ D, {\bf 48}, 4835 (1993).
\bibitem{7} B. Carter, Int.\ J.\ Theo.\ Phys., {\bf 42}, 1317
(2003).
\bibitem{8} R. Cartas-Fuentevilla, Phys.\ Lett.\ B, {\bf 563}, 107 (2003).
\bibitem{9} C. Crncovi\'c and E. Witten, in {\it Three Hundred Years
of Gravitation}, edited by S. W. Hawking and W. Israel (Cambridge
University Press. Cambridge, 1987).
\bibitem{10} R. Cartas-Fuentevilla, {\it Conformal symmetry of the
phase space formulation for topological string actions}, preprint
(2004).
\end{thebibliography}
\end{document}